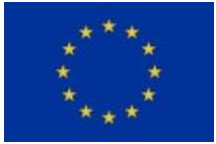
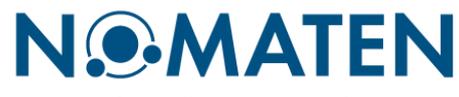

This work was carried out in whole or in part within the framework of the NOMATEN Centre of Excellence, supported from the European Union Horizon 2020 research and innovation program (Grant Agreement No. 857470) and from the European Regional Development Fund via the Foundation for Polish Science International Research Agenda PLUS program (Grant No. MAB PLUS/2018/8), and the Ministry of Science and Higher Education's initiative "Support for the Activities of Centers of Excellence Established in Poland under the Horizon 2020 Program" (agreement no. MEiN/2023/DIR/3795).

# Dissociative Mechanism from NH3 and CH4 on Ni-Doped Graphene: Tuning Electronic and Optical Properties

A. Aligayev[a,b,c], U. Jabbarli[d], U. Samadova[e,f], F. J. Dominguez–Gutierrez[c], S. Papanikolaou[c], Qing Huang[a,b]

[a]Key Laboratory of High Magnetic Field and Iron Beam Physical Biology, Institute of Intelligent Machines, Hefei Institutes of Physical Science, Chinese Academy of Sciences, Hefei 230031, China

[b]Science Island Branch of Graduate School, University of Science and Technology of China, Hefei, 230026, China

[c]NOMATEN Centre of Excellence, National Centre for Nuclear Research, ul. A. Sołtana 7, 05-400 Otwock, Poland

[d]Faculty of Physics, University of Warsaw, Pasteura 5, PL-02093 Warsaw, Poland

[e]University of Electronic Science and Technology of China, Chengdu, 05-400, PR.China

[f]Institute of Physics Ministry of Science and Education Republic of Azerbaijan, H.Javid 131, AZ1143

**Abstract**

In this study, we employ a multi-scale computational modeling approach, combining density functional theory (DFT) and self-consistent charge density functional tight binding (SCC-DFTB), to investigate hydrogen ($H_2$) production and dissociation mechanisms from ammonia ($NH_3$) and methane ($CH_4$) on pristine and nickel-doped graphene. These two-dimensional materials hold significant potential for applications in advanced gas sensing and catalysis. Our analysis reveals that Ni-doped graphene, validated through work function calculations, is a promising material for gas separation and hydrogen production. The samples with adsorbed molecules are characterized by calculating chemical potential, chemical hardness, electronegativity, electrophilicity, vibrational frequencies, adsorbtion and Gibbs energies by DFT calculations. Methane molecules preferentially adsorb at the hexagonal ring centers of graphene, while ammonia inter- acts more strongly with carbon atoms, highlighting distinct molecular doping mechanisms for $CH_4$ and $NH_3$. Dynamic simulations show that $CH_4$ splits into $CH_3$+H, with Ni-doped graphene facilitating enhanced hydrogen transmission, while $NH_3$ dissociates into $NH_2$+H, which may lead to $N_2H_4$ formation. Our non-equilibrium Green's function (NEGF) simulations demonstrate increased H-atom transmission on Ni-doped graphene during gas interactions. These findings suggest that Ni-doped graphene is superior to pristine graphene for applications in gas separation, hydrogen production, and high-sensitivity sensors.

**Introduction**

The overuse of fossil fuels has not only led to their scarcity but also caused environmental damage due to the emission of greenhouse gases. Thus, hydrogen is emerging as an efficient, sustainable and clean energy source because of its high gravimetric energy content (120 MJ kg$^{-1}$), which is significantly higher than that of gasoline, and its emissions free of CO$_x$ [1, 2]. This makes hydrogen particularly attractive for fuel cell applications in energy generation. However, hydrogen's storage and distribution pose challenges due to its low volumetric energy density (8.96 GJ m$^{-3}$ in liquid form) and its boiling point of -253 °C, making large-scale operations expensive and potentially unsafe [3, 4, 5]. To address these issues, hydro- gen can be chemically stored in other compounds, such as methanol (CH$_3$OH), methane (CH$_4$), ammonia (NH$_3$) and its derivatives, and ammonia borane (e.g. NH$_3$BH$_3$), that is easier to transport and store, and then released through a chemical reaction [6, 7, 8, 9]. Among these potential hydrogen carrier compounds, ammonia–NH$_3$ stands out due to its high hydrogen con- tent (17.6 wt%), high energy density (13.6 GJ m$^{-3}$), ease of liquefaction under mild conditions (0.86 MPa at 20 ◦C), a narrow flammability range in air (16-25 vol%) compared to H$_2$ (4-75 vol%), and established industrial production methods [10, 11, 12, 13, 14, 15]. To make ammonia a viable source for H$_2$ production, its decomposition must occur at temperatures below 150 °C [16]. Creating catalysts that can effectively operate under these conditions is particularly difficult due to the reaction's endothermic nature. Furthermore, at low temperatures, the rate-limiting step is the desorption of nitrogen, which can lead to catalyst poisoning from strongly adsorbed nitrogen atoms.

Metal active sites must therefore bind nitrogen strongly enough to hold ammonia molecules but not so strongly that nitrogen desorption is hindered after the N-H bonds are broken [17]. Thus, the metal-nitrogen binding energy often serves as a critical factor in designing catalysts for ammonia decomposition. Currently, thermocatalytic decomposition of ammonia or ammonia cracking (2NH$_3$ ↔ 3H$_2$ + N$_2$; ∆H = 46.22 kJ mol$^{-1}$) using catalysts like ruthenium (Ru) and nickel (Ni) is a common dehydrogenation technique, actively studied and refined worldwide [18, 13]. Li et al. used ruthenium catalysts onto–on different carbon substrates to study their effectiveness in decomposing NH$_3$ for hydrogen production [19, 20]. They found that when active sites dispersed uniformly throughout surfaces of graphite, the Ru/graphite catalyst achieved NH$_3$ conversion rates as high as 95% at 550 °C under a gas hourly space velocity (GHSV) of 30,000 mL g$^{-1}$h$^{-1}$. Ni-based catalysts have desired availability, economically advantageous due to their high catalytic activity, making them a compelling choice for ammonia decomposition at low temperature [21, 22, 23]. Unfortunately, a single component of Ni catalysts does not suffice to achieve efficient activity and stability during the ammonia cracking process. Consequently, there is a strong preference for catalyst systems that integrate nickel with supporting materials characterized by robust thermal stability, extensive surface area, and effective electron transfer conductivity, all aimed at enhancing NH$_3$ decomposition performance. In this regard, carbon-based materials have proven to be excellent catalytic supports in ammonia cracking processes [24] due to their superior conductivity, which facilitates electron transfer to the nickel active sites, and their outstanding surface area.

Graphene, a two-dimensional multifunctional monolayer is composed of carbon atoms arranged in a tightly-packed honeycomb lattice [25]. It is a material of significant interest and challenge, with diverse applications across electronics, semiconductors, fuel cells, and gas storage. The charge carriers in graphene, which behave as massless Dirac fermions, endow it with remarkable electrical capabilities, including high carrier mobility (200,000 cm$^2$ V$^{-1}$ s$^{-1}$) at 300 K [26]. Beyond its exceptional electronic properties, graphene's high thermal, chemical, and mechanical stability, along with its large surface area to volume ratio (2630 m$^2$ g$^{-1}$), exceptional conductivity (106 S cm$^{-1}$), and excellent optical transmittance (~97.7%) [27, 28] make it an ideal 2D support layer for metallic nanoparticles in chemical sensor and heterogeneous catalysis [29, 30, 31].

In this study, we have applied SCC-DFTB and DFT computational techniques to perform an extensive analysis [32, 33] of the effectiveness of Ni-doped graphene in chemical sensors and its role in NH$_3$ and CH$_4$ decomposition, our study is structured as follows: In Section 3 the numerical simulation specifics are delineated, including a brief explanation of the SCC-DFTB approach and the computational techniques employed. Dissociative mechanisms of methane and ammonia, optical adsorbance, and electron transport calculations are presented as a function of absortion rates. Finally, in section 4, we provide concluding remarks.

**2. Computational methods**

The first-principles density functional theory calculations were performed within the Vienna ab initio simulation package (VASP) [34, 35]. The generalized gradient approximation (GGA) framework with the Perdew- Burke-Ernzerhof (PBE) [36] functional was employed to describe exchange-correlation functional. Projector augmented wave (PAW) [37] potentials were used for the electron ion interaction. The plane–wave basis cutoff energy was set to 400 eV, which has been determined to be well converged. A 12 × 12 × 1 Monkhorst-Pack [38] k–point mesh was employed in all calculations for Brillouin zone integration. The convergence criteria were set to 0.05 eV/Å$^{-1}$ for forces and 1×10$^{-5}$ eV for the total energy. To prevent potential interactions between adjacent structures in the z-direction, a vacuum layer of approximately 20 Å was introduced. All geometric structures were depicted by using the OVITO software [39]. The Self-Consistent-Charge Density-Functional Tight- Binding (SCC-DFTB) approach, an approximation of density functional theory (DFT) utilizing second-order charge corrections, as implemented in the DFTB+ software [40], facilitates large-scale simulations with comparative accuracy and efficiency, often surpassing traditional DFT methods. Originating from a simplification of Kohn–Sham DFT, SCC-DFTB adopts a tight-binding framework to characterize Hamiltonian eigenstates using atomic-like basis sets, substituting the Hamiltonian with a parameterized matrix dependent solely on internuclear distances and orbital symmetries. Hence, the ground state density is described as perturbed by density fluctuations: $\rho(r) = \rho_0(r) + \delta\rho(r)$, where the total energy of the system can be expressed as [40]:

$$EDFTB[\rho_0 + \delta\rho] = E_0[\rho_0] + E_1[\rho_0, \delta\rho] + E_2[\rho_0, (\delta\rho)^2] + E_3[\rho_0, (\delta\rho)^3], \quad (1)$$

with $E_0[\rho_0]$ and $E_1[\rho_0, \delta\rho]$ approximation take into account only valence minimal basis set of the system within a linear combination of atomic orbitals (LCAO) ansatz; and $E_2[\rho_0, (\delta\rho)^2]$ and $E_3[\rho_0, (\delta\rho)^3]$ includes information about the Mulliken charges and describe the change of the chemical hardness of an atom computed from DFT.

*2.1. Optical absorbance and electron transport*

Optical absorption is explored within DFTB frame- work, treating it as an electronic dynamic process in response to an external electric field [41, 42]. The conventional adiabatic approximation is employed to determine the time evolution of the electron density matrix, achieved through the time integration of the Liouville–von Newmann equation.

$$i\frac{\partial \hat{\rho}}{\partial t} = S^{-1}\hat{H}\hat{\rho} - \hat{\rho}S^{-1} \quad (2)$$

where $\hat{\rho}$ is the single electron density matrix, $S^{-1}$ is the over- lap matrix, and $\hat{H}$ is the system Hamiltonian that includes the external electric field as $\hat{H} = \hat{H}_0 + E_0\delta(t - t_0)\hat{e}$ with $E_0$, the magnitude of the electric field, and $\hat{e}$, its direction. Within the context of linear response, the absorbance I(ω) is determined as the imaginary part of the Fourier transform of the induced dipole moment induced by an external field. In our investigation, the external field strength was fixed at (E0 = 0.1) V/Å. The induced dipole moment was computed over a time span of (200) fs (with a time step of $\Delta t = 0.01$ fs). To mitigate noise, the Fourier transform was executed with an exponential damping function, utilizing a damping constant of (5) fs.

We calculate the electron transmission probability using the non-equilibrium Green's function (NEGF) method, which generalizes the Landauer approach to compute the tunneling current between two conducting leads subject to open boundary conditions, as implemented in the DFTB code [43, 44]. The NEGF approach is also used to consistently calculate charge densities by solving the Pois- son equation under non-equilibrium conditions. In Fig. 1, we present a detailed illustration of the geometric con- figuration of the graphene structure, highlighting the specific regions involved in the electron transport calculations. The system is designed with an extended device region and two semi-infinite bulk contacts, each defined by their Principal Layers (PLs), also known as drain-source layers. PLs consist of atoms that interact only with adjacent PLs, allowing for an accurate description of bulk systems through the iterative Green's function solver algorithm. The drain section represents the region where electrons exit the device, while the source section corresponds to the region where electrons enter. To study the impact of adsorbed molecules on electron transport, we placed substances such as $NH_3$ and $CH_4$ in the device region. This setup allows for precise capture of interactions between the adsorbed molecules and the graphene surface, facilitating an accurate analysis of their effects on the transmission probability.

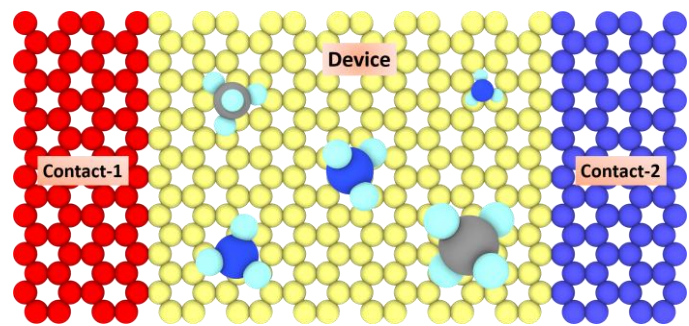

Figure 1: (Color on-line). Schematic representation of an optimized graphene monolayer structure used for electron transport probability calculations. The red and blue regions represent the contact layer 1 and 2, respectively. $NH_3$ and $CH_4$ molecules are present in the device region, illustrating their interaction with the graphene surface.

*2.2. Molecular beam simulations*

By utilizing a reactive molecular beam with high kinetic energy, gas-surface reactions involving high activation barriers can be accelerated. This approach allows for overcoming the barriers that would otherwise require elevated pressures when using a random gas with a Maxwell- Boltzmann distribution [45]. In order to investigate the absorption dynamics and potential dissociation of $CH_4$ and $NH_3$ molecules interacting with the pristine and Ni– doped graphene sheets as close as possible to molecular beam experiments, we employed semi–classical molecular dynamics simulations in a $8 \times 8 \times 1$ supercell. The surface was optimized and equilibrated to 300 K using a Nose–Hoover thermostat [46]. For the adsorption dynamics, we defined a target area of 1 $nm^2$ on the surface and randomly distributed the molecules using the velocity Verlet algorithm [46, 32]. The molecules were emitted vertically with random orientations from an initial distance of 0.7 nm above the surface, with an impact energy of 2 eV. We generated 1000 independent trajectories for each molecule, employing a time step of 0.1 fs. The simulations are performed for 350 fs, a duration carefully selected to ensure convergence while allowing the molecules to move away from the carbon sheets while remaining bonded.

This time frame effectively balanced the prevention of detachment and convergence in our molecular dynamics simulations. We have previously employed this approach to study the hydrogenation mechanisms of fullerene cages and 2D carbon sheets [47], the optoelectronic properties of graphenylene [48], and the dynamic physisorption pathways of molecules on alumina surfaces [49], demonstrating excellent agreement with first-principles DFT calculations. Our methods consider that an approaching atomic ion would be neutralized by the electron Fermi sea of the conducting sheet where the total electronic energy of the adiabatic system is solved quantum-mechanically at each time step; an electronic temperature of 1000 K, equivalent to Fermi–Dirac smearing, is used in the MD simulations.

## 3. Results and discussion

The work function, a critical factor affecting device performance, is an essential characteristic of nanomaterials that can be considered [50, 51]. Therefore, we computed the work function by considering a $7 \times 7 \times 1$ 2D- graphene supercell using DFT from a unit cell of 2 carbon atoms with an internuclear distance of 1.42 Å and lattice vectors arranged to reproduce a hexagonal sheet [48]. In order to prevent periodic image interactions, 15 Å inter– slab vacuum was applied along the (001) direction and periodic boundary conditions on the x and y axes. Then, the work function is calculated as:

$$\Phi = E_{vac} - E_F, \quad (3)$$

where $E_{vac}$ and $E_F$ represent the energy of a stable electron within the vacuum level and the Fermi level, respectively. As shown in Fig. 2, the calculated work function of 4.42 eV for 2D-graphene, which shows reasonable agreement with the experimental value of 4.56 eV [51]. The black solid line corresponds to the experimental work function of graphene, serving as a benchmark. The red dashed line represents the calculated work function for pristine graphene, which, although close to the experimental value, slightly underestimates it. This discrepancy may stem from inherent approximations in density functional theory (DFT), including the choice of exchange–correlation functional, which can affect the accuracy of the calculated electronic properties. The blue dashed line illustrates the electrostatic potential for Ni–doped graphene, where a single carbon atom in the $7 \times 7 \times 1$ graphene is substituted by a nickel atom, to keep a low concentration of Ni. The introduction of Ni as a dopant results in a slight reduction in the work function compared to pristine graphene. This reduction indicates that Ni doping modifies the electronic structure of graphene, offering potential for tuning the material's electronic properties. The observed trends in work function modification due to Ni doping align with the expected influence of metal atoms on graphene's electronic environment, suggesting that Ni–doped graphene could be a promising candidate for applications in electronic and optoelectronic devices [31].

In addition, we compute the work function for the surfaces with the molecules noticing that adsorption of $CH_4$ on graphene lowered the work function to 4.27 eV, suggesting that a methane molecule may donate electrons to the graphene surface, thereby enhancing its reactivity. Conversely, $NH_3$ adsorption on graphene raised the work function to 4.39 eV, indicating that an amonnia molecule likely redistributes electrons, resulting in an increased surface potential. For Ni–doped graphene, $CH_4$ adsorption further reduced the work function to 4.24 eV, suggesting a stronger electron donation from the molecule to the sur- face. In contrast, NH3 adsorption r resulted in a work function of 4.49 eV, which is lower than the work function of $NH_3$–adsorbed pristine graphene but higher than that of Ni–doped graphene alone. This indicates a less pronounced interaction between $NH_3$ and Ni–doped graphene in comparison to $CH_4$.

We present the density of states (DOS) for pristine splitting suggests the possibility of spin polarization in the system, as transition metal doping often leads to spin-orbit coupling and magnetic effects. These changes in the band structure imply that transition metal-doped graphene could exhibit tailored electronic, magnetic, and catalytic properties, making it a promising candidate for energy- related applications such as hydrogen evolution reactions (HER) and oxygen reduction reactions (OER) [55].

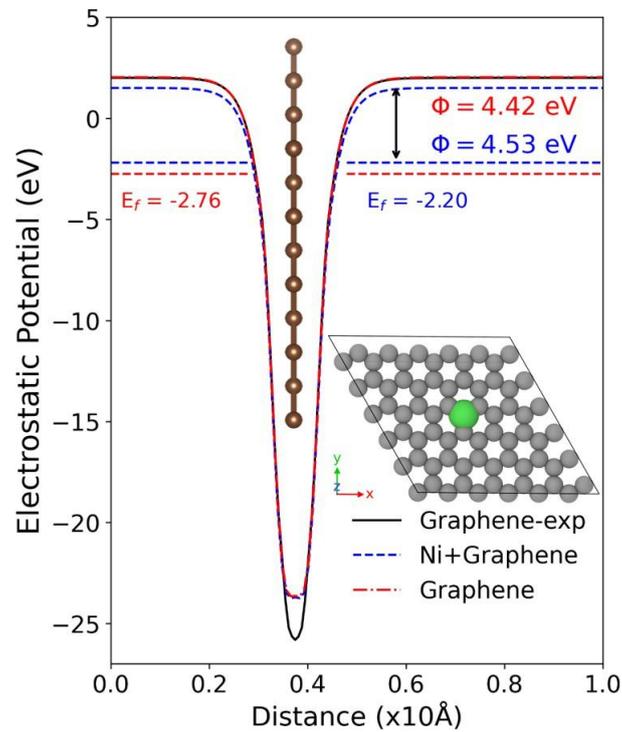

Figure 2: (Color on-line) Planar average of the potential energy across single layer pristine (red dashed-dot line), Ni-doped (blue dashed line), and experimental reference (black continuous line) [51, 52] are given together with DFT results. We add the visualization of the $7 \times 7 \times 1$ Ni–doped graphene sheet as reference.

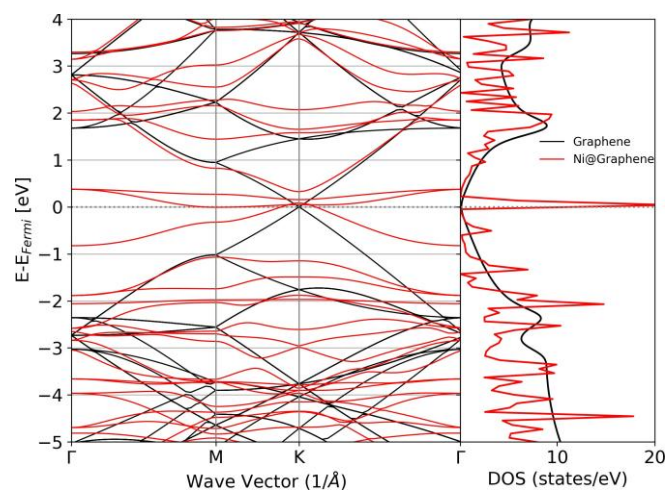

Figure 3: (Color on-line) Dos for a) pristine and b) Ni-doped graphene for $7 \times 7 \times 1$ supercell, as shown in Fig 2.

*3.1. Physisorption pathways*

The interaction potentials between $NH_3$ and $CH_4$ with pristine and Ni-doped graphene are investigated using adiabatic calculations to determine potential energy curves, where only self-consistent calculations are performed for the molecules and fully relaxed periodic sheets at various distances and adsorption sites, allowing the atoms in the molecule and sheet to relax while keeping the molecule- surface distance fixed; accounting for van der Waals in- graphene in Fig. 3, by DFT calculations. In pristine graphene, tractions through dispersion corrections [56]. Then, the Fermi energy ($E_F$) aligns with the Dirac energy ($E_D$), confirming that the intrinsic band gap of graphene is zero, as reported in literature. While, the substitution of a single Ni atom in the graphene sheet introduces an impurity band, resulting in the appearance of several small density of states peaks. The 3d-orbitals of Nickel hybridize with the graphene π–orbitals [53], leading to the formation of new electronic states near the Fermi level. Ni doping, which typically introduces additional electrons, causes an upward shift of the Fermi level from the Dirac point.

In addition, we compare the band structures of pristine and Ni–doped graphene, obtained through DFT calculations, analysing their behavior along the high–symmetry points in the Brillouin zone (Γ-M-K-Γ). The black solid–lines represent the band structure of pristine graphene, which exhibits the well–known linear dispersion near the K–point, corresponding to the Dirac cone at the Fermi energy [54]. This gapless structure is a signature of graphene's semi–metallic nature, where conduction and valence bands meet at a single point. The band structure of Ni–doped graphene is presented by red dotted line resulting in additional electronic states near the Fermi level, which are attributed to the hybridization between Ni 3d orbitals and graphene's pz orbitals. This hybridization alters the density of states near the Fermi level, potentially introducing a localized magnetic moment and altering the conductive properties of the material. The band flattening seen in the Ni-doped graphene is indicative of increased electron localization and reduced group velocity, which could have implications for transport properties, such as reduced mobility and enhanced scattering. Furthermore, the band total energies, denoted as $E_{(z)}$, pertain to the system of molecule–2D material at separations z ranging from 0.5 to 8.5 Å above the surface. This span helps define the computation of the adsorption potential concerning the distance separation. The total energy is then computed as:

$$E(z) = E_{Tot} - (E_{2D} + E_{Molecule}) , \quad (4)$$

The notations used in the calculations are: $E_{2D}$, representing the total energy of the 2D material; $E_{Molecule}$, signifying the total energy of the isolated molecule types: $NH_3$ and $CH_4$; and $E_{Tot}$, referring to the system's energy at each z-distance. The binding energy, denoted as $E_b$, is derived from $E_{(zmin)}$, with $z_{min}$ denoting the equilibrium distance between the molecule and the surface. The computation involves evaluating the total energy for the system of molecule–2D material evaluating 3 unique adsorption sites such as: top, bridge and hole accounting for the unit cell of graphene. While adjusting the distance be- tween the surface and the molecules' center of mass along the z-axis. For the Ni–doped graphene, an extra adsorption site located above the Ni atom was included in the calculations.

In Fig. 4, we present physisorption pathways of the $CH_4$ molecule on pristine and Ni–doped graphene, showing the minimum–energy adsorption site selected from all candidates. The methane molecule belongs to the tetrahedral point group (Td), which leads us to consider different orientations relative to the 2D materials: (i) three hydrogen atoms bonded to the carbon atom interacting with the adsorption site, (ii) two hydrogen atoms positioned above the site bonded to carbon, and (iii) one hydrogen atom directly interacting with the adsorption site. The results from DFT (VASP) and SCC-DFTB (DFTB+) calculations show close agreement across these configurations. In the first configuration being the most stable, the $CH_4$ molecule achieves a bond length of 2.5 Å at adsorption site located within the hexagonal ring, attributed to 2p bonding between C atoms. Similar behaviour was observed for the Ni–doped graphene, due to the weak Ni-C bond respect to C-H one. This configuration is presented in the inset figure for the pristine graphene case, for visualization. This agreement across methods validates the modeling approach for further simulations of methane ad- sorption by the 2D materials.

The $NH_3$ molecule belongs to the C3v symmetry group, characterized by a C3 principal axis of rotation and three vertical planes of symmetry, which leads us to consider similar configurations as in the methane case. The molecule is rotated to explore configurations where (i) all three hydrogen atoms interact with the adsorption sites, (ii) two hydrogen atoms interact, and (iii) a single hydrogen atom interacts. The physisorption pathway for the configuration with the lowest energy is presented in Fig. 5, characterized by small binding energies and large bonding distances. The good agreement between the two approaches DFT and SCC–DFTB can also validate the modeling for the case of ammonia where it is noteworthy that the physisorption process, particularly when all three hydrogen atoms are in contact with the adsorption site at the center of the hexagonal ring, is relatively straightforward to manage.

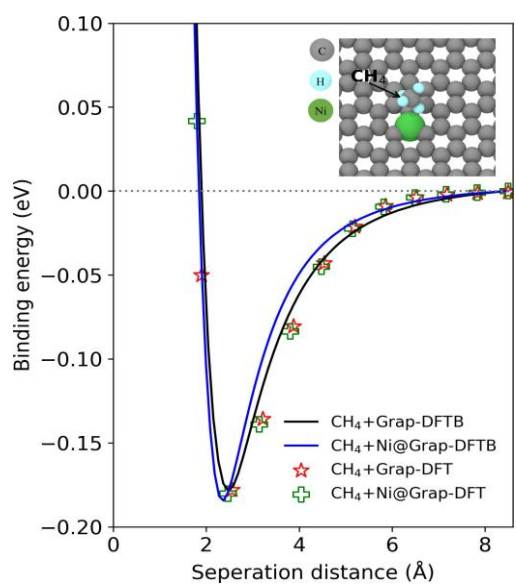

Figure 4: (Color on-line) The PEC curve represents the bonding length between $CH_4$ molecule and pristine and Ni–doped graphene. In the inset figure, we show the most stable configuration with the minimal energy at the hole of the hexagonal ring.

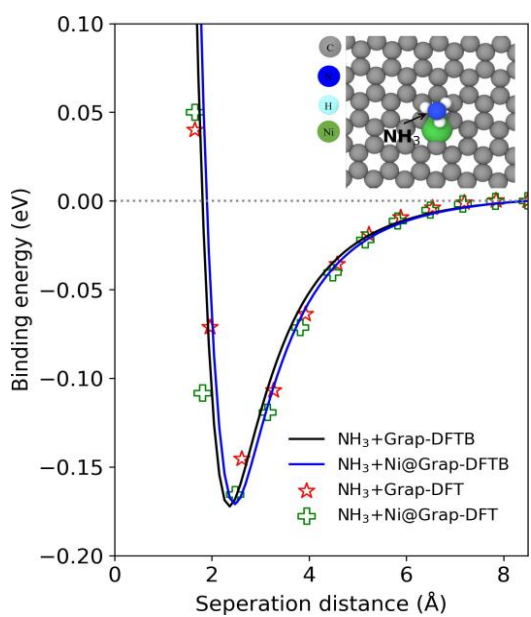

Figure 5: (Color on-line) Potential energy curves showing the bond length between the $NH_3$ molecule and both pristine and Ni-doped graphene. The inset displays the ammonia molecule at the adsorption site within the hexagonal ring on both surfaces.

This process is expected to be completely reversible due to the low binding energy and rapid sorption kinetics for pristine and Ni–doped sheets. In addition, the physisorption energy is quite low for all systems, indicating that very high pressures are required to achieve significant surface coverage at room temperature. At 150°C, even higher pressures are necessary. For a potential well depth of 0.17 eV, at 1 bar and 150°C, the equilibrium coverage could be estimated to be around 0.002 monolayers (ML). To achieve 20 % surface coverage, a pressure of approximately 100 bar would be required. This information is important for our investigation of dissociative mechanisms and provides an insight to an experimental linkage from our modeling.

Once the binding energy between the molecules and graphene sheets is determined, the next step involves optimizing the system's energy by positioning the $NH_3$ and $CH_4$ molecules at the identified adsorption sites without any distance constraints letting the system explore additional degrees of freedom e.g., lateral shifts, tilting, and rotation of the molecule relative to the surface. The optimization was done by using the SCC–DFTB approach with conjugate gradient method, the structures were fully relaxed with respect to the volume, shape, and internal atomic positions until the atomic forces were less than $10^{-4}$ eV/Å for the whole numerical cell. In Fig. 6, we present the results for $CH_4$ and $NH_3$ molecules adsorbed on pristine and Ni-doped graphene. For the pristine case (Fig. 6a-b), the distance between the carbon atom of $CH_4$ and the nearest carbon atom of the pristine graphene sheet is 3.36 Å. In the Ni–doped system, the distance is 4.70 Å to the closest carbon atom and 6.07 Å to the Ni atom. Notably, the repulsion between the $NH_3$ molecule and the Ni atom causes the graphene sheet to bend, altering its electronic structure. For the $NH_3$ molecule adsorbed on pristine graphene (Fig. 6c), the distance between nitro- gen and the nearest carbon atom is 3.12 Å. The hydrogen bond lengths are slightly modified due to interaction with graphene, causing a minor rotation of the molecule. In the Ni-doped case (Fig. 6d), the Ni atom is attracted to the $NH_3$ molecule, bending the sheet upwards, which may impact the material's electronic properties. The Ni-N distance is 3.0 Å, while the Ni-C distance is 1.02 Å due to the stronger C-H bond compared to the C-Ni bond. These observations are consistent with other Ni-doped graphene systems [57].

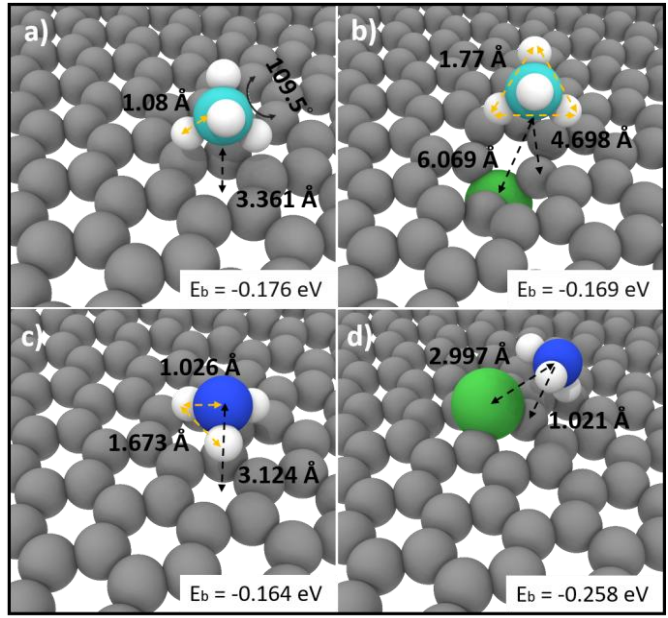

Figure 6: Optimized systems for graphene and Ni-doped graphene considering CH$_4$ and NH$_3$ molecules at the binding energy positions from SCC-DFTB calculations.

To further investigate the effects of NH$_3$ and CH$_4$ interactions with pristine and Ni-doped graphene on the electronic properties, we present charge difference visualizations in Fig. 7. Panels a-b) show the pristine graphene results, while panels c-d) depict the Ni-doped graphene with the respective molecules. The presence of Ni significantly alters the sp2 bonding and adsorption behavior of both molecules compared to pristine graphene.

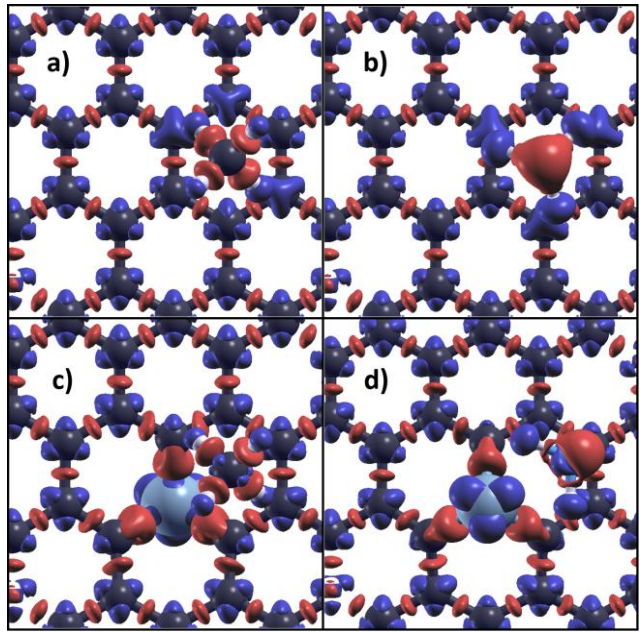

Figure 7: Charge difference distribution for graphene with a) NH$_3$, b) CH$_4$, c) NH$_3$ on Ni-doped graphene, and d) CH$_4$ on Ni@graphene. The charge distribution is shown in red for negative and blue for positive regions. The presence of Ni alters the electronic structure of graphene, facilitating the bonding of hydrogen atoms to carbon atoms.

For NH$_3$, a pronounced charge polarization is observed upon adsorption on Ni-doped graphene, indicating an enhanced interaction likely driven by charge transfer and orbital hybridization between the nitrogen lone pair and Ni d-orbitals.

| System | $\mu$ | $\eta$ | $\chi$ | $\omega$ |
|---|---|---|---|---|
| Pristine | 2.42 | 0.27 | -2.24 | 9.41 |
| Graphene + NH$_3$ | 2.38 | 0.28 | -2.38 | 10.14 |
| Graphene + CH$_4$ | 2.03 | 0.24 | -2.03 | 8.79 |
| Ni@Graphene + NH$_3$ | 2.67 | 0.29 | -2.24 | 8.65 |
| Ni@Graphene + CH$_4$ | 2.53 | 0.33 | -3.03 | 8.79 |

Table1. Chemicalpotential ($\mu$), chemicalhardness($\eta$), electronegativity($\chi$), and electrophilicity($\omega$) of pristine and Ni-doped graphene before andafteradsorptionofNH3 and CH4 in eV.

This suggests that NH3 undergoes stronger chemisorption on the doped surface, facilitated by the catalytic activity of Ni, which activates adsorption sites and induces electronic rearrangements. In contrast, CH$_4$ shows only minor charge difference variations, consistent with weak physisorption, though slightly enhanced by Ni doping. The interaction between CH$_4$ and Ni-doped graphene is dominated by van der Waals forces with minimal charge transfer. These results suggest that Ni doping enhances graphene's reactivity, especially towards polar molecules like NH$_3$, while having a more limited effect on nonpolar molecules like CH$_4$.

*3.2. Characterization of optimized surfaces*

To star the characterization of the systems, in Tab 1 we display the chemical potential ($\mu$), chemical hardness ($\eta$), electronegativity ($\chi$), and electrophilicity ($\omega$), of pristine and Ni-doped graphene, emphasizing the effects of doping and adsorbate interactions. Pristine graphene shows relatively high stability and moderate reactivity, while interactions with NH$_3$ and CH$_4$ reduce its chemical potential and hardness slightly, indicating mild reactivity. Ni doping significantly enhances graphene's reactivity, as reflected in higher chemical potential and lower hardness, especially in interactions with NH$_3$. Lower electronegativity values for Ni-doped graphene demonstrate its in- creased capacity to interact with electron-rich molecules. To provide more information about the optimized systems, we conducted Reduced Density Gradient (RDG) calculations to investigate the interactions of both $CH_4$ and $NH_3$ with Ni–doped and pristine graphene. RDG is calculated using the electron density and its first derivative as:

$$RDG(r) = \frac{1}{2*(3\pi^2)^{4/3}} * \frac{|\nabla\rho(r)|}{\rho(r)^{4/3}}$$

used to describe the derivation from a homogeneous electron distribution, and computed by using NCIPLOT4 software [58]. The RDG plot for Ni–doped graphene interacting with $CH_4$ (Fig. 8) reveals a distinctive distribution pat- tern, with pronounced blue and green regions that suggest stronger non-covalent interactions. This enhanced inter- action profile indicates that Ni doping significantly amplifies the interaction strength with $CH_4$, highlighting Ni–doped graphene's potential in applications requiring elevated adsorption and interaction efficiency. These results underscore the effectiveness of Ni-doped graphene in facilitating interactions with $CH_4$, reinforcing its relevance for practical applications.

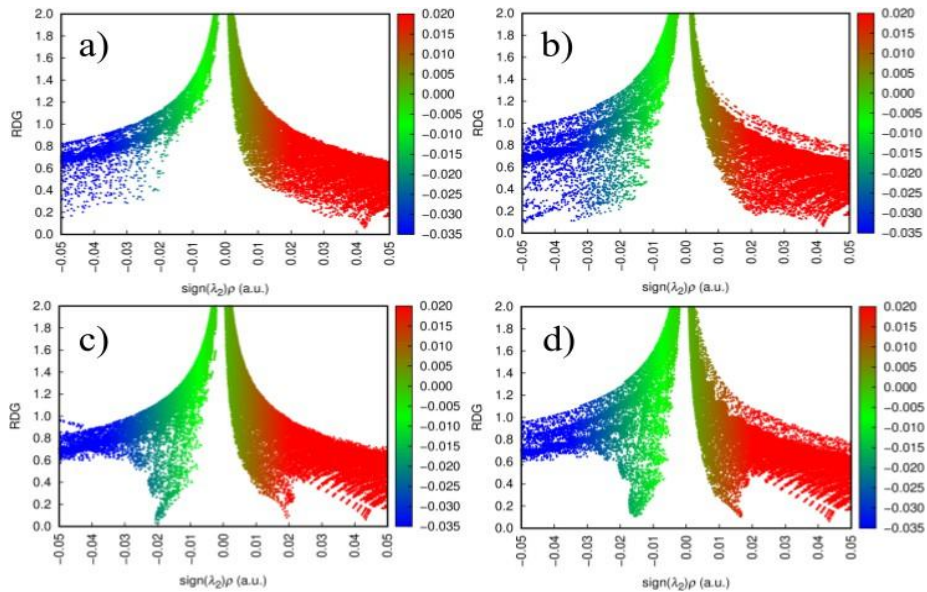

Figure 8: Reduced Density Gradient for a) graphene+$CH_4$, b) graphene+$NH_3$, c) Ni@graphene+$CH_4$ and Ni@graphene+$NH_3$.

*3.3. Optical Adsorption*

In Fig. 9, we present the normalized optical absorption spectrum of pristine graphene (a) and Ni-doped graphene (b) calculations were using the Liouville-von Neumann equation for methane and ammonia molecules. The optical absorption calculations were performed on unconstrained, optimized systems guided by the physisorption pathways, simplifying the computational process for more efficient and accurate determination of atomic configurations. We noticed that the optical absorbance observed in both pristine and Ni-doped graphene with $NH_3$, compared to $CH_4$, can be influenced by several factors: i) the introduction of Ni dopants modifies the electronic structure of graphene, as shown in Fig. 3; ii) the electron-donating nature of $NH_3$ and $CH_4$ induces charge transfer effects, possibly shifting the Fermi level and altering the available electronic states (as shown in Fig. 8). In addition, $NH_3$ may also promote surface functionalization, introducing nitrogen-containing groups that generate new energy states conducive to modifying the optical properties. The combined effects of electron donation, charge transfer, and potential structural changes likely contribute to the observed optical absorbance behavior of Ni-doped graphene with $NH_3$ and $CH_4$.

*3.4. Electron Transport*

The chemical molecular doping, where physical adsorption of $CH_4$ and $NH_3$ molecules on graphene surface modifies the electronic properties of graphene based material, can create an effective offset voltage that will lead to an additional displacement field for band gap opening. In order to carry out an analysis of the effects of molecule doping in graphene sheets, we decide to consider a armchair ribbon where the band gap is dependent on the size of the ribbon, providing the opportunity to tailor electronic and quantum transport properties while considering chemical doping substituting a C atom by a Ni one and molecular doping simultaneous. In Fig. 10 a) we report calculations for the electron transmission to investigate the impact of Ni doping and molecular adsorption on the electronic transport properties of the 2D material. For pristine graphene, we notice a high transmission near the Fermi level, indicative of its superior electronic conductivity, as expected from the density of states for a ribbon as shown in the SI (see Fig S8). However, Ni-doped graphene (red dotted curve) shows minor modifications in the transmission near the Fermi level, suggesting the presence of localized states without a significant drop in transmission or formation of a large bandgap. The orientation-dependent variations in transmission for $NH_3$ adsorption on both pristine and Ni-doped graphene underscore the interplay between molecular orientation and electronic structure modifications. These findings underscore the role of doping and molecular interactions in tailoring the electronic properties of graphene, offering valuable insights for the design of graphene-based electronic and sensing devices.

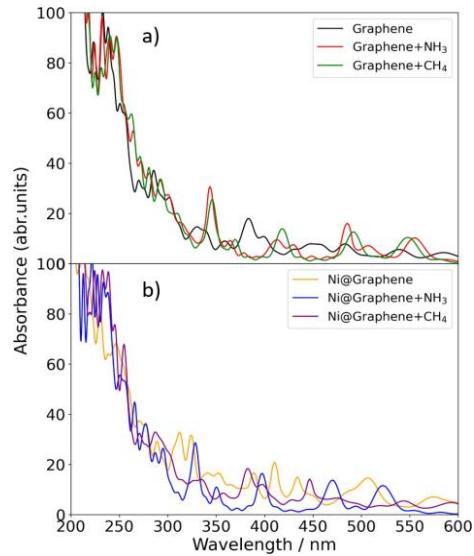

Figure 9: (Color on-line) Optical absorbance spectra of pristine graphene (a) and Ni-doped graphene (b) in the visible range, with $NH_3$ and $CH_4$ molecules adsorbed on the surfaces.

The adsorption of $NH_3$ and $CH_4$ further modulates the electron transport, with $NH_3$ showing a more substantial reduction in transmission probability compared to $CH_4$, reflecting stronger interaction and scattering effects.

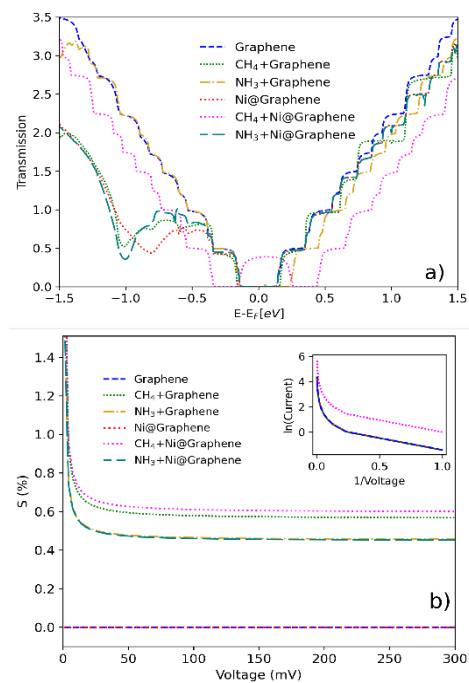

Figure 10: Total transmission summed over all channels for pristine and Ni-doped graphene nanoribbons with chemical molecular doping, shown in a). The reported results highlight size effects on the band gap value. Sensitivity of the 2D sheets is demonstrated in b) with different adsorbed molecules. The inset displays the linear current-voltage (I- V) characteristics in the range of 0-300 mV.

Fig 10 b) shows the difference of the current for each molecule X as a function of the voltage as:

$$S = 100\% \frac{|I_X - I_g|}{I_X}$$

with IX of each molecule and the surface and Ig the current of the pristine and Ni-doped graphene. The tunneling currents for the sheets as a function of the voltage for various adsorbed molecules are shown in the inset plots by I-U characteristics graphs for voltages below 300 mV. Our results demonstrate a sensitivity increase of 0.5 % for ammonia and 0.65% for methane upon molecular doping, with Ni–doped graphene and methane showing the highest sensitivity and a pronounced effect on the current-voltage response. This enhanced sensitivity is attributed to charge polarization induced by Ni doping, which alters the electronic properties of the graphene sheet, and the sp2-hybridization of the carbon atoms upon methane adsorption.

*3.5. Dissociative mechanisms*

The analysis of dissociative chemisorption mechanisms for methane and ammonia provides valuable insights into the key energetic and dynamical aspects of a prototypical polyatomic–surface reaction, which also serves as the rate–limiting step in the industrially significant steam reforming process. Dissociative chemisorption is a well–known type of surface reaction where a molecule adsorbs onto a solid surface and subsequently breaks into two or more fragments. This process is crucial in many catalytic reactions, particularly in heterogeneous catalysis, where the surface of a solid catalyst interacts with gas or liquid- phase molecules. In this work, the modeling of these interactions is based on semi–classical molecular dynamics approach.

In Tab 2 compares experimental and calculated values for Zero-Point Energy (ZPE), Thermal Entropy (T*S), and Vibrational Frequency for $NH_3$ and $CH_4$ molecules in various configurations. Furthermore, we calculate the ad- sorption energy, $E_{Ads}$, and Gibbs energy, EGibbs, as a reference. The DFT–calculated values for isolated $NH_3$ and $CH_4$ show close alignment with experimental data, indicating the accuracy of our computational approach. For pristine and Ni-doped graphene, both ZPE and T*S values are negligible, reflecting the planar structure and symmetry of graphene, which limits out-of-plane vibrational contributions.

Table 2: Zero–Point Energy (ZPE), Thermal Entropy (T*S), and Vibrational Frequency values for $NH_3$ and $CH_4$ adsorbed on pristine and Ni- doped graphene, alongside experimental data [59]. the adsorption energy, $E_{Ads}$, and Gibbs energy, $E_{Gibbs}$, are reported as reference.

| System | ZPE (eV) | T*S (eV) | Vib. Freq. (THz) | $E_{Ads.}$ (eV) | $E_{Gibbs}$ |
|---|---|---|---|---|---|
| $NH_3$ (Exp.) | 0.8944 | 0.5956 | 1.1602 | - | - |
| $NH_3$ (DFT) | 0.9574 | 0.4753 | 1.0849 | - | 1.43 |
| $CH_4$ (Exp.) | 1.1753 | 0.5759 | 1.0881 | - | - |
| $CH_4$ (DFT) | 1.1864 | 0.4984 | 1.0475 | - | 1.68 |
| Pristine Graphene | 0.0 | 0.0 | - | - | - |
| Ni-doped Graphene | 0.0 | 0.0 | - | - | - |
| Graphene + $CH_4$ | 1.2418 | 0.06812 | 1.0542 | -0.33 | 1.31 |
| Graphene + $NH_3$ | 0.9699 | 0.1607 | 0.9987 | -0.38 | 1.13 |
| Ni-doped Graphene + $CH_4$ | 1.2645 | 0.0892 | 1.7681 | -0.39 | 1.35 |
| Ni-doped Graphene + $NH_3$ | 0.9851 | 0.1654 | 1.1587 | -0.45 | 1.15 |

Adsorption of $NH_3$ and $CH_4$ on graphene and Ni-doped graphene shows increased ZPE and T*S, suggesting enhanced interactions, especially in Ni-doped configurations. Thermal energy corrections at 298.15 K for $NH_3$ and $CH_4$ were computed as approximately 1.07 kcal/mol and 0.89 kcal/mol, respectively, due to contributions from vibrational, rotational, and translational degrees of freedom, with $NH_3$ showing a higher correction due to its dipole and complex vibrational modes.

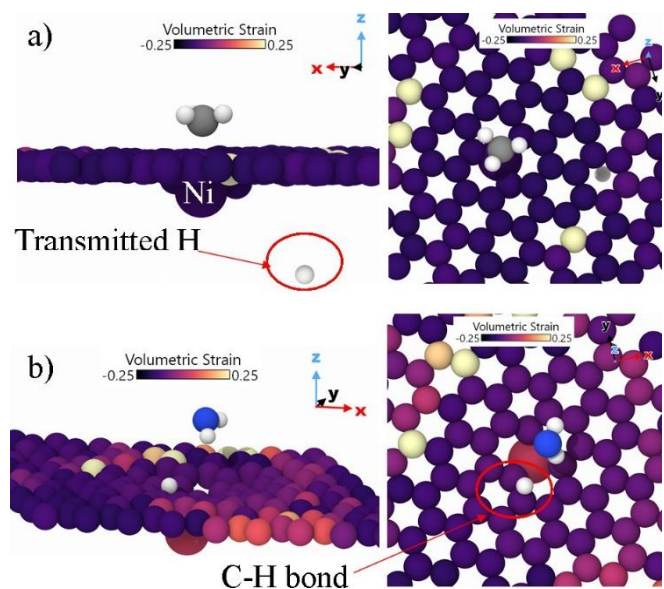

Figure 11: Dissociative mechanisms of $CH_4$ (a) and $NH_3$ (b) following collisions with Ni-doped graphene, based on a selected MD simulation from a set of multiple trials. For methane, the presence of Ni in graphene is observed to promote the transmission of H atoms, a behavior not seen with pristine graphene. For both 2D materials, the formation of azanyl ($NH_2$) molecules is observed, with hydrogen fragments showing a high probability of bonding to carbon atoms in the graphene. Volumetric strain is shown to visualize the effect of the collision on the 2D material.

In the context of methane dissociation mechanisms on pristine graphene, the $CH_4$ molecule approaches the 2D material, and as it gets closer, the C-H bond breaks, leading either to adsorption of the resulting fragments onto a C atom of the graphene or to the emission of a hydrogen atom. The outcome depends on the kinetic energy of the methane molecule, resulting in the formation of $CH_3$ radicals, as reported in our previous work [48]. For Ni-doped graphene, the dissociative mechanism of $CH_4$ molecules following a collision with the doped surface at the site of minimal energy, selected from a set of MD simulations at different adsorption sites, is illustrated in Fig. 11a. In this case, the kinetic energy of the $CH_4$ molecule is transferred to the 2D material, resulting in an increase in volumetric strain around the Ni impurity. Although the dissociation mechanism is similar to that observed for pristine graphene, the presence of Ni alters the process, allowing for the transmission of the hydrogen atom depending on the kinetic energy of the incoming molecule. At an impact energy of 2 eV, the morphology of the 2D sample is distorted without breaking the Ni-C or C-C bonds, but these bonds may break at higher impact energies. Here, the volumetric strain calculated by OVITO [39] shows the region where the methane molecule collides with the 2D material. The dissociative mechanisms of $NH_3$ are of significant interest in the context of dehydrogenation processes using either pristine or Ni- doped graphene [60]. Based on our MD simulations, we observed that the $NH_3$ molecule dissociates into $NH_2$ and H upon collision with the 2D material. This mechanism allows the hydrogen atom to bond to a carbon atom on the graphene surface, regardless of the presence of a Ni impurity, as shown in Fig. 11b). It is also observed that $NH_2$ can bond to the carbon atoms of the sheet, forming an N-C single bond. At a kinetic energy of 5 eV, the dissociation of ammonia leads to hydrogen atoms attaching to the 2D material, while the $NH_2$ molecule is reflected. This observation also explains the formation mechanisms of $N_2H_4$ molecules in experiments where $NH_2$ radicals are present, leading to significant production of hydrazine ($N_2H_4$) as a reactive intermediate. The collision of $NH_3$ molecules with the 2D material is observed to affect more carbon atoms compared to the methane case, as indicated by the volumetric strain analysis. Finally, in order to calculate the recovery time for sensing mechanisms of the pristine and Ni-doped graphene by DFT calculations, which primarily describes the detachment or desorption of a molecule already adsorbed on the surface from the obseved events in our MD simulations. We first identify the TS by CI-NEB (Climbing Image Nudged Elastic Band) method with the VTST (Variable Temperature String) tool [61], which allowed us to refine the reaction pathway with greater accuracy. After identifying an approximate TS, we further refined it through transition state optimization, confirming it as a first-order saddle point by verifying the presence of a single imaginary frequency along the reaction coordinate. Next, we obtained the activation energies as $E_{activation} = E_{TS} - E_{reactant}$ where the transition to $CH_3 + H$ requires a high energy barrier of 4.1 eV on pristine graphene, but this is reduced to 2.9 eV on Ni-doped graphene, indicating enhanced catalytic activity. While, $NH_3$ into $NH_2+H$, which has an energy barrier of 0.38 eV on pristine graphene. However, on Ni-doped graphene, this energy barrier is slightly reduced to 0.36 eV, indicating enhanced catalytic efficiency. In the supplementary material, we present results for the complete set of TS for each molecule. The next step is to calculate the recovery time by using the Arrhenius Equation:

$$\tau = \frac{1}{\nu} \exp\left(\frac{E_a}{K_B T}\right)$$

where the $\nu$ is obtained from the vibrational frequency, the $E_a$ is the activation energy. Thus, from our calculations we noticed that the recovery times are quite long which indicates a very stable adsorption, suggesting that desorption (or recovery) would be practically unobservable under standard conditions at room temperature.

## 4. Concluding remarks

In this study, we investigate the effects of chemical and Ni-doping on the electronic and optical properties of graphene using ab-initio (DFT) and density functional tight binding (DFTB) simulations. By calculating the work function, we validate Ni-doped graphene as a promising material for applications in gas separation and hydrogen production. Our band structure and density of states (DOS) calculations reveal band flattening in Ni-doped graphene, indicative of increased electron localization and reduced group velocity, which could impact transport properties, such as mobility and scattering. Potential energy curves calculations show that methane molecules preferentially adsorb at the hexagonal ring center of graphene, while ammonia molecules tend to interact more with carbon atoms. This highlights distinct molecular doping mechanisms for $CH_4$ and $NH_3$. Dynamic simulations reveal that $CH_4$ splits into $CH_3+H$ on both pristine and Ni-doped graphene, with H atoms transmitting more readily on Ni-doped sheets. Similarly, $NH_3$ dissociates into $NH_2+H$, which can lead to the formation of $N_2H_4$ in experiments, with $NH_2$ molecules showing higher adsorption on Ni-doped graphene.

In conclusion, while pristine graphene shows limited H-atom transmission during $CH_4$ interaction, Ni-doping significantly enhances this effect. Our findings indicate a measurable difference in the adsorption energies of $NH_3$ and $CH_4$ on Ni-doped graphene sheets, suggesting potential for selective detection through distinct adsorption- based mechanisms. Although the energy difference is modest, it can significantly influence adsorption-induced electronic properties such as conductivity or resistance, which are commonly leveraged to identify specific gases in sensor applications. By calibrating the sensor to detect these subtle variations in electronic response, selective identification of $NH_3$ and $CH_4$ can be achieved. Furthermore, the variation in adsorption energy is likely to impact charge transfer dynamics at the gas-surface interface. $NH_3$, with relatively higher interaction energy, may induce greater charge redistribution on the surface than $CH_4$, leading to a stronger and more detectable electrical signal. This characteristic serves as an additional metric for selective sensing, as the enhanced charge redistribution effect of $NH_3$ can result in a more pronounced sensor response, as demonstrated by our optical absorption and current-voltage measurements. Overall, Ni-doped graphene-based materials show promise over pristine graphene for applications in gas separation, hydrogen production, and high-speed sensors.

**Acknowledgments**

Research was funded through the European Union Horizon 2020 research and innovation program under Grant Agreement No. 857470 and from the European Regional Development Fund under the program of the Foundation for Polish Science International Research Agenda PLUS, Grant No. MAB PLUS/2018/8, and the initiative of the Ministry of Science and Higher Education "Support for the activities of Centers of Excellence established in Poland under the Horizon 2020 program" under Agreement No. MEiN/2023/DIR/3795. We gratefully acknowledge Polish high-performance computing infrastructure PLGrid (HPC Center: ACK Cyfronet AGH) for providing computer facilities and support within computational Grant No. PLG/2024/017084